\documentclass{appolb}
\usepackage{graphicx}
\usepackage{hyperref}

\begin{document}
\title{Evidence for top quark production in nucleus-nucleus collisions with the CMS experiment
\thanks{Presented at Quark Matter 2022 - XXIXth international conference on ultrarelativistic
nucleus-nucleus collisions}%
}
\author{L. Alcerro
\thanks{On behalf of the CMS Collaboration}
\thanks{Supported by the Nuclear Physics program\href{https://pamspublic.science.energy.gov/WebPAMSExternal/Interface/Common/ViewPublicAbstract.aspx?rv=00d4fe0f-48a0-4d4a-baf1-c70867d9e499&rtc=24&PRoleId=10}{ DE-FG02-96ER40981} of the U.S. Department of Energy.}
\address{The University of Kansas}
{ 
\address{}
}
\address{}
}
\maketitle
\begin{abstract}
Evidence for the production of top quarks in heavy ion collisions is reported in a data sample of lead-lead collisions recorded in 2018 by the CMS experiment at a nucleon-nucleon center-of-mass energy of $\sqrt{s_{_{\mathrm{NN}}}} =$ 5.02 TeV, corresponding to an integrated luminosity of $1.7\pm0.1\,\mathrm{nb}^{-1}$. Top quark pair ($\mathrm{t\bar{t}}$) production is measured in events with two opposite-sign high-$p_\mathrm{T}$ isolated leptons ($\ell^\pm\ell^\mp =\,\mathrm{e}^{+} \mathrm{e}^{-},\,\mu^{+} \mu^{-},\,\mathrm{and}\,\mathrm{e}^{\pm} \mu^{\mp}$). We test the sensitivity to the $\mathrm{t\bar{t}}$ signal process by requiring or not the additional presence of b-tagged jets, and hence demonstrate the feasibility to identify top quark decay products irrespective of interacting with the medium (bottom quarks) or not (leptonically decaying W bosons). To that end, the inclusive cross section ($\sigma_\mathrm{t\bar{t}}$) is derived from likelihood fits to a multivariate discriminator, which includes different leptonic kinematic variables with and without the b-tagged jet multiplicity information. The observed (expected) significance of the $\mathrm{t\bar{t}}$ signal against the background-only hypothesis is 4.0 (5.8) and 3.8 (4.8) standard deviations, respectively, for the fits with and without the b-jet multiplicity input. After event reconstruction and background subtraction, the extracted cross sections are $\sigma_\mathrm{t\bar{t}} = 2.03 ^{+0.71}_{-0.64}$ and $2.54 ^{+0.84}_{-0.74}\,\mu\mathrm{b}$, respectively, which are lower than, but still compatible with, the expectations from scaled proton-proton data as well as from perturbative quantum chromodynamics predictions. This measurement constitutes the first crucial step towards using the top quark as a novel tool for probing strongly interacting matter.
\end{abstract}
  
\section{Introduction}
The top quark is the heaviest particle in the standard model (SM). With a mass of roughly $173$ GeV and a near unity Yukawa coupling, it decays before hadronization. At LHC energies, top quarks are primarily produced in $\mathrm{t\bar{t}}$ pairs by gluon fusion and decay most of time in a b quark and a W boson. \\
Top quark production is important to understand various aspects of QCD. In proton-proton collisions top quark production is useful to constrain proton PDFs and a key process to compute parameters such as the $|V_\mathrm{tb}|$ matrix element of the CKM matrix (see e.g. \cite{CMS:2020vac,CMS:2021vhb}). Moreover, proton-proton collisions serve as baseline to collisions of heavy nuclei.  In nucleus-nucleus collisions, $\mathrm{t\bar{t}}$ production is important to test nuclear PDFs whereas it sets the scenario for studies on the QGP time evolution. \\
Unlike other jet quenching probes, the top quark has the particularity that, depending on its momentum, it can decay before or within the QGP. Taking ``snapshots" at different times (or momentum) and with sufficient statistics, one would be able to resolve the time evolution of the QGP. Phenomenological studies of hadronically decaying W bosons coming from $\mathrm{t\bar{t}}$ quark pairs in a QCD medium show the potential of top quarks to provide insights of the time structure of the QGP \cite{Apolinario:2017sob}.

\section{$\mathrm{t\bar{t}}$ in PbPb at $\sqrt{s_{_{\mathrm{NN}}}} = 5.02 \,\mathrm{TeV}$}
\begin{figure}
\begin{center}
\includegraphics[width=\textwidth]{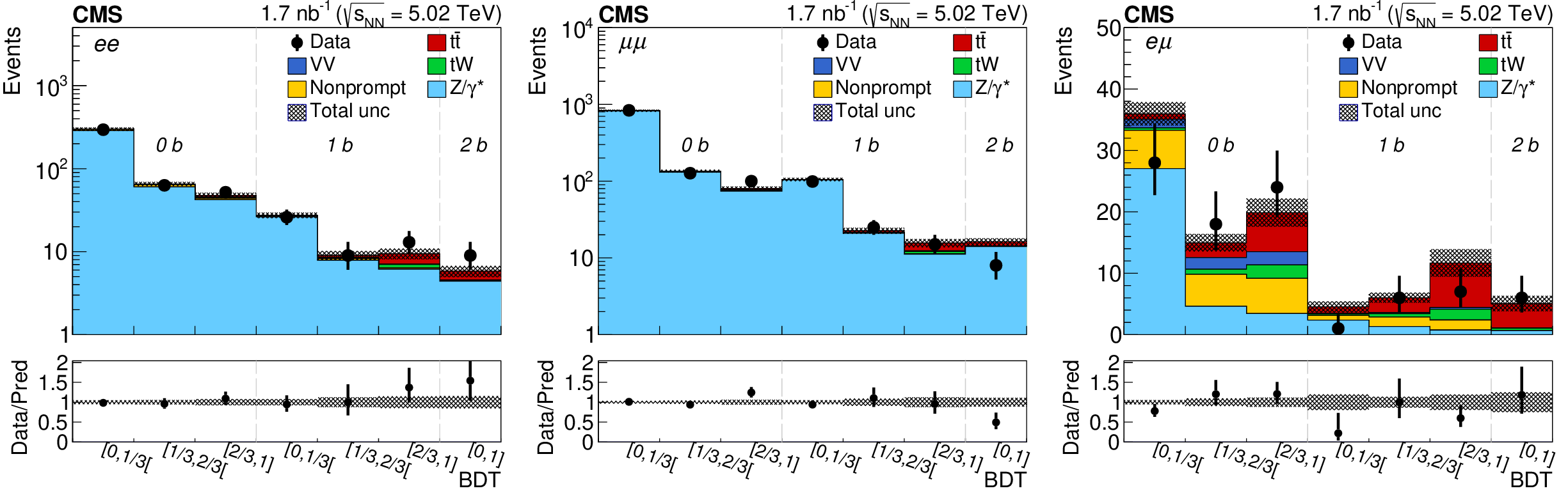}
\caption{BDT discriminator distributions in the $e^+e^-$ (left), $\mu^+ \mu^-$ (middle), and $e^{\pm} \mu^{\mp}$ (right) final states separately for the 0b-, 1b-, and 2b-tagged jet multiplicity categories. Plots taken from Ref. \cite{CMS:2020aem}.}
\label{pbpb_2}
\end{center}
\end{figure}

\begin{figure}
\begin{center}
\includegraphics[width=0.49\textwidth]{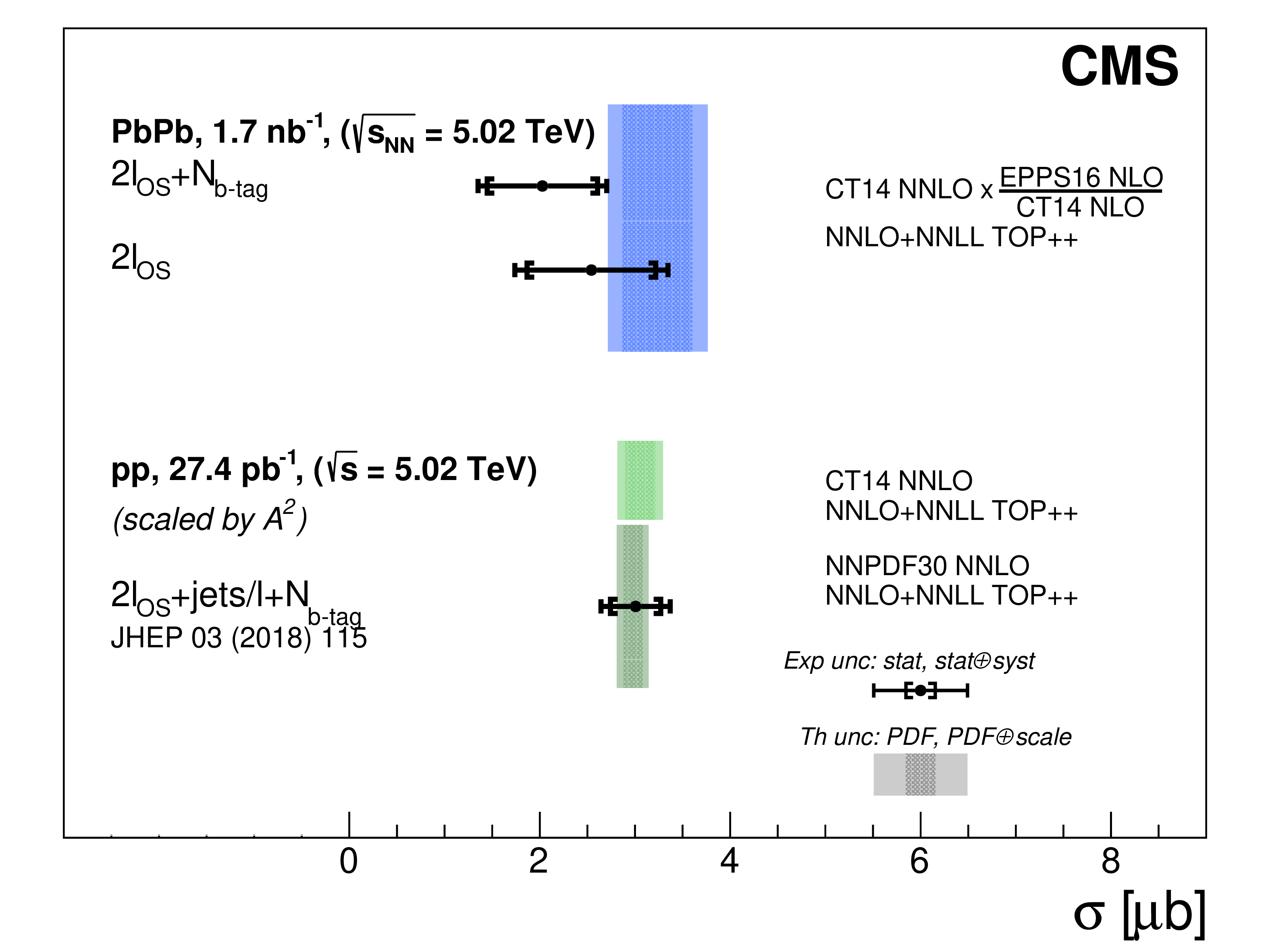}
\includegraphics[width=0.49\textwidth]{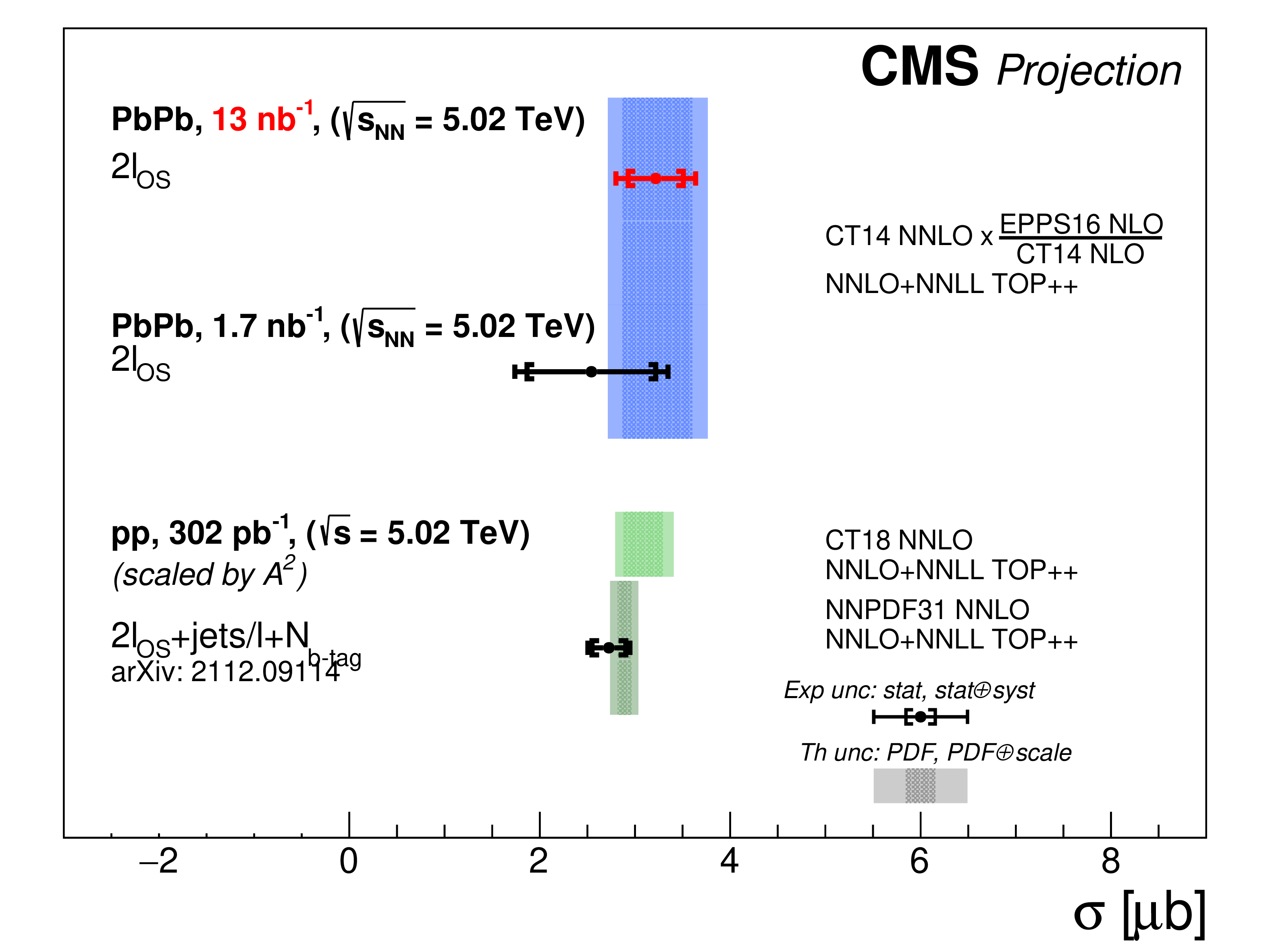}
\caption{Inclusive $\mathrm{t\bar{t}}$ cross sections measured with two methods in the combined $e^+e^-$, $\mu^+\mu^-$, and $e^{\pm} \mu^{\mp}$ final states in PbPb collisions at $\sqrt{s_{_{\mathrm{NN}}}} = 5.02 \,\mathrm{TeV}$, and pp results at the same energy (left). Projection for $\mathrm{t\bar{t}}$ in PbPb at HL-LHC (right) \cite{CMS:2020aem}.}
\label{pbpb_3}
\end{center}
\end{figure}

The very first evidence of top quark production in heavy nuclei was achieved by the CMS Collaboration \cite{CMS:2008xjf} using data recorded in 2018, corresponding to an integrated luminosity of $1.7\, \mathrm{nb}^{-1}$ of lead-lead (PbPb) collisions \cite{CMS:2022bjp} at nucleon-nucleon center-of-mass energy of $\sqrt{s_{_{\mathrm{NN}}}} =$ 5.02 TeV \cite{CMS:2020aem}. \\
The dilepton channel is exploited in this analysis with and without inclusion of information coming from b tagged jets. Data is required to contain two opposite sign (OS) leptons with $p_\mathrm{T}>25$ (20) GeV and $|\eta|<2.1$ (2.4) for electrons (muons) with no nearby hadronic activity. The presence of b-tagged jets is further exploited in a second method to enhance the signal. Jets coming from b quarks are tagged using a combined secondary vertex algorithm \cite{CMS:2017wtu} relying on machine learning techniques.  \\
In both methods, a boosted decision tree (BDT) method is trained to discriminate genuine leptons with high $p_\mathrm{T}$ between signal and background processes (see Fig. \ref{pbpb_2}). In order to minimize effects of the imprecise knowledge of the jet properties in the heavy ion environment, BDTs use kinematic properties only. Likelihood fits to binned BDT distributions are performed separately for the two methods to extract the cross section, obtaining $\sigma_\mathrm{t\bar{t}} = 2.03 ^{+0.71}_{-0.64}\,\mu\mathrm{b}$
with the $2\ell_{\mathrm{OS}} +$ b jets method and 
$2.54 ^{+0.84}_{-0.74}\,\mu\mathrm{b}$ 
with $2\ell_{\mathrm{OS}}$. The results are compatible with theoretical calculations and pp at 5.02 TeV scaled by the number of binary nucleon-nucleon collisions in PbPb as well, as is shown in Fig. \ref{pbpb_3} (left).
\section{$\mathrm{t\bar{t}}$ in pp at $\sqrt{s}= 5.02\,\mathrm{TeV}$}
The first measurement of the total $\mathrm{t\bar{t}}$ cross section at 5.02 TeV in the $\ell+$ jets and dilepton channels was performed by the CMS Collaboration using data recorded in November 2015 \cite{CMS:2017zpm} with a data sample that corresponds to an integrated luminosity of 27.4 pb$^{-1}$ \cite{CMS:2016zvd}. \\
This result has been recently updated with an increased  integrated luminosity \cite{CMS:2021nvp} of more
than an order of magnitude compared to the data set previously mentioned \cite{CMS:2021gwv}. This analysis takes into account the dilepton channel only, which in combination with the $\ell+$jets result from 2015 data \cite{CMS:2017zpm}, gives an updated cross section measurement of $\sigma_{\mathrm{t\bar{t}}} = 63.0 \pm 4.1 \textrm{  (stat)}\pm 3.0 \textrm{ (syst + lumi)}$  pb. 
\section{Projection at HL-LHC}
Projection of $\mathrm{t\bar{t}}$ in the High-Luminosity LHC era assuming a PbPb integrated luminosity of 13 nb$^{-1}$ is shown in Fig. \ref{pbpb_3} (right). This projection (red) takes into account only dilepton final states (no b jets) and is compared with the measured (black) $\mathrm{t\bar{t}}$ cross section, scaled pp data discussed in the previous section, and theoretical predictions at NNLO+NNLL accuracy in QCD. This projection shows that total uncertainties are expected to be halved with respect to Run 2 uncertainties. 
\section{Summary}
In summary, the CMS Collaboration has presented evidence of top quark production in PbPb collisions at nucleon-nucleon center of mass energy of 5.02 TeV and integrated luminosity of $1.7 \pm 0.1\,\mathrm{nb}^{-1}$. Two methods were used for the cross section extraction: one using kinematic properties of the final dilepton pair only and other including b jet identification techniques, both in agreement with theory. This result represents the first step for detailed studies using top quarks in nuclear interactions in view of higher luminosities where, as we show for the HL-LHC case, uncertainties are expected to be halved in comparison to Run 2 data. 

\bibliographystyle{auto_generated.bst} 
\bibliography{Alcerro.bib}
\end{document}